\documentclass[superscriptaddress,showkeys,reprint,aps,pre]{revtex4-1}
\usepackage{blindtext}
\usepackage{graphicx}
\usepackage{mathtools}
\usepackage{amsmath}
\usepackage{amssymb}
\usepackage{amsthm}
\usepackage{amsfonts}
\usepackage{braket}
\usepackage{multirow}
\usepackage{makecell}
\usepackage{soul}
\usepackage[export]{adjustbox}

\usepackage{xcolor}

\begin{document}
\title{Random logic networks: from classical Boolean to quantum dynamics }
\author{Lucas Kluge}
\affiliation{Potsdam Insitute for Climate Impact Research$,$ Telegrafenberg$,$ 14473 Potsdam$,$ Germany}
\affiliation{Institute of Physics and Astronomy$,$ University of Potsdam$,$ Karl-Liebknecht-Str. 24/25$,$ 14476$,$ Potsdam$,$ Germany}
\email{kluge@pik-potsdam.de}
\author{Joshua E.~S.~Socolar}
\affiliation{Department of Physics$,$ Duke University$,$ Durham$,$ NC$,$ 27708$,$ USA}
\author{Eckehard Schöll}
\affiliation{Institut  für  Theoretische  Physik$,$  Technische  Universität  Berlin$,$  Hardenbergstr$.$ 36$,$  10623  Berlin$,$  Germany}
\affiliation{Potsdam Insitute for Climate Impact Research$,$ Telegrafenberg$,$ 14473 Potsdam$,$ Germany}

\begin{abstract}
We investigate dynamical properties of a quantum generalization of classical reversible Boolean networks.  The state of each node is encoded as a single qubit, and classical Boolean logic operations are supplemented by controlled bit-flip and Hadamard operations.  We consider synchronous updating schemes in which each qubit is updated at each step based on stored values of the qubits from the previous step.  We investigate the periodic or quasiperiodic behavior of quantum networks, and we analyze the propagation of single site perturbations through the quantum networks with input degree one.  A non-classical mechanism for perturbation propagation leads to substantially different evolution of the Hamming distance between the original and perturbed states.
\end{abstract}

\maketitle
\section{Introduction}
Random Boolean networks 
exhibit behaviors that lend insights into
a variety of fields, serving as generic models describing the dynamics of complex systems ranging from neural~\cite{rosin2013control}, social~\cite{hurford2001random}, protein interaction~\cite{kauffman2003random}, or game theoretic networks~\cite{alexander2003random}. Studies of random Boolean networks began in earnest with Kauffman's introduction of a model framework for gene regulatory networks, which consisted of a set of $N$ nodes, each having $K$ random inputs from nodes chosen at random~\cite{kauffman1969metabolic}. At each time step, each node is updated according to a randomly assigned Boolean logic operation on its $K$ inputs.  

The number of possible states of a Boolean network is finite (equal to $2^N$). Thus for any initial condition and any deterministic sequence chosen for updating the nodes, the network eventually settles on a periodic attractor. The nature of the set of such attractors for large networks has been a major research topic~\cite{flyvbjerg1988exact,Nielsen:2011:QCQ:1972505,drossel2005number,Derrida_1986,socolar2003scaling,samuelsson2003superpolynomial,greil2005dynamics,drossel2005number,samuelsson2006exhaustive,kaufman2005properties,drossel2005number,paul2006properties}, with special attention devoted to a dynamical phase transition that occurs as either $K$ or the probabilities of assigning different Boolean functions are varied.

The present work is motivated by the possibility of observing a dynamical phase transition or qualitatively different dynamics in random networks of quantum mechanical gates, where the classical states of the nodes are generalized to qubit states and the set of Boolean operations is expanded to include quantum logic functions.
Random Boolean networks generally involve high rates of dissipation, as multiple different combinations of the input states to a given node produce the same output state.
A na\"ive introduction of quantum logic into the system retains a high level of dissipation associated with
the information loss intrinsic to the projection operations required to create a state of a qubit that is independent of its previous state. Preliminary studies revealed that the introduction of intrinsically quantum mechanical operations leads to a reduction of the dynamics to a single stable fixed point. Removing the projection operators from the quantum system results in reversible dynamics whose classical analogue is found in the reversible Boolean networks introduced by Coppersmith et al.~\cite{coppersmith2001reversible,coppersmith2001reversibleII}.
%
%\lk{We use the term dissipation to indicate a loss of information. If all information of a system is conserved, the system is time reversible.}
%
%\lk{Before considering reversible networks, we briefly investigated the impact of non-classical operation on the classical Kauffman network. To realize the classical Kauffman network as a quantum circuit, we used projection operations. Projection operations map to a single state, independently of the previous state. As a result, the system loses information. When introducing the Hadamard operation, which is a non-classical operation that we will discuss in depth later on, the systems showed the same dynamics independently of the wiring. More specifically the system only showed a single stable attractor without any further dynamics. We decided that in order to maintain the effects introduced by non-classical operations, we need not consider reversible networks. Since our problems were the non-unitary projection operations, we decided to remove them. As a results we obtain a reversible Boolean network that was previously investigated by Coppersmith et al.~\cite{coppersmith2001reversible,coppersmith2001reversibleII} .}
%Coppersmith et al.introduced a class of Boolean networks that exhibit reversible dynamics, which by definition does not involve the erasure of information and hence produces no dissipation.  They noted that
%
In contrast to dissipative systems, every state of a reversible Boolean network lies on a periodic orbit; there are no transients.  
The key questions then concern the distribution of periods (cycle lengths), the scaling of the number of periodic orbits with $N$ for different choices $K$ or the set of Boolean functions employed, and the stability of cycles under single qubit perturbations.

In the present paper, we consider a modification of  reversible random Boolean networks that incorporates inherently quantum operations at some or all of the nodes.  We will refer to these as ``quantum networks.''
A quantum network consists of a set of qubits together with a set of operations determining how each qubit is updated based on the states of a subset of qubits referred to as its inputs.  Like classical random Boolean networks, these quantum networks are updated in discrete time steps.  Unlike their classical counterparts, however, the state space of the system is not necessarily discrete, as a given qubit can be in an arbitrary superposition of its two basis states (which we take to be the classical Boolean states).  
%As applications of qubit networks, discrete-time quantum spin networks have recently been discussed~\cite{sakurai2021chimera}.

In a quantum network, the operations analogous to Boolean logic functions are unitary operations on the states of some set of qubits.  Classical reversible Boolean operations are a subset of these, and a minimal extension of this set includes quantum operations on single qubits.  A typical operation might consist, for example, of a classical (reversible) operation applied to several qubits, followed by a quantum operation that creates a superposition of the 0 and 1 states of the output (e.g., a Hadamard operation). 
To extend the results of Coppermith et al.\ to the quantum realm, we formulate the reversible dynamics in a way that allows for a natural insertion of quantum operations.  Each node in the Coppersmith model is now represented by {\em two} qubits, one representing the current state of that node and the other representing its state one time step in the past.  This enables an efficient formulation of the dynamics in which the future state of the network depends only on the present state. In a recent study, Franco et al.\ introduced an equivalent formulation of the class of reversible quantum networks and examined the behavior of networks with $N \leq 6$ \cite{franco2021random}.

Our goal is to identify any new effects that arise when unitary operations are included that lead the system out of the discrete classical state space.  
We focus here on the 
cycles that arise in small networks and on the divergence of trajectories that differ initially in the state of a single qubit.
To study the latter, we extend the concept of classical Hamming distance to one that is well suited to describing a special set of quantum networks employing only Hadamard operations and controlled-not gates. We study the divergence of trajectories in networks in which each qubit receives information from a single other qubit, and describe qualitative differences between the quantum and classical cases.

\section{Quantum generalizations of reversible Boolean networks}
To establish notation, we introduce an example classical reversible Boolean network and describe the formalism for representing it as a quantum circuit.  We then describe a particular extension of the set of Boolean truth functions to intrinsically quantum mechanical operations, which will produce the quantum dynamics we are interested in studying.

Consider the reversible classical Boolean networks introduced by Coppersmith et al.~\cite{coppersmith2001reversible,coppersmith2001reversibleII}. A network consists of $N$ spin variables which can be in state $s = -1$ or $+1$. (Note: We use $s$ here rather than $\sigma$ to avoid confusion with Pauli $\sigma$ matrices below.) The time evolution of each of the Boolean variables depends on the values of $K$ other variables. The network updating rule is given by:
\begin{equation}
    s_{t+1}^{(i)} =  s_{t-1}^{(i)}F^{(i)}\left({\bf S}^{(i)}_t\right),
    \label{eq:reversible}
\end{equation}
where $s_t^{(i)}$ is the value of spin $i$ at discrete time step $t$, ${\bf S}^{(i)}_t$ represents the state vector consisting of the $K$ input spins to node $i$ at time $t$, and $F^{(i)}$ is a Boolean function of $K$ inputs. Each node is updated using its own value at time $t-1$ and the output of a truth function $F^{(i)}$. Because $s_t^{(i)}$ can take only the values $\pm 1$,
this model can be written in the equivalent form
\begin{equation}
    s_{t+1}^{(i)} s_{t-1}^{(i)}=F^{(i)}({\bf S}^{(i)}_t),
    \label{eq:reversible2}
\end{equation}
which makes manifest the time-reversal invariance of the dynamics.

\begin{figure}
\includegraphics[width=\linewidth]{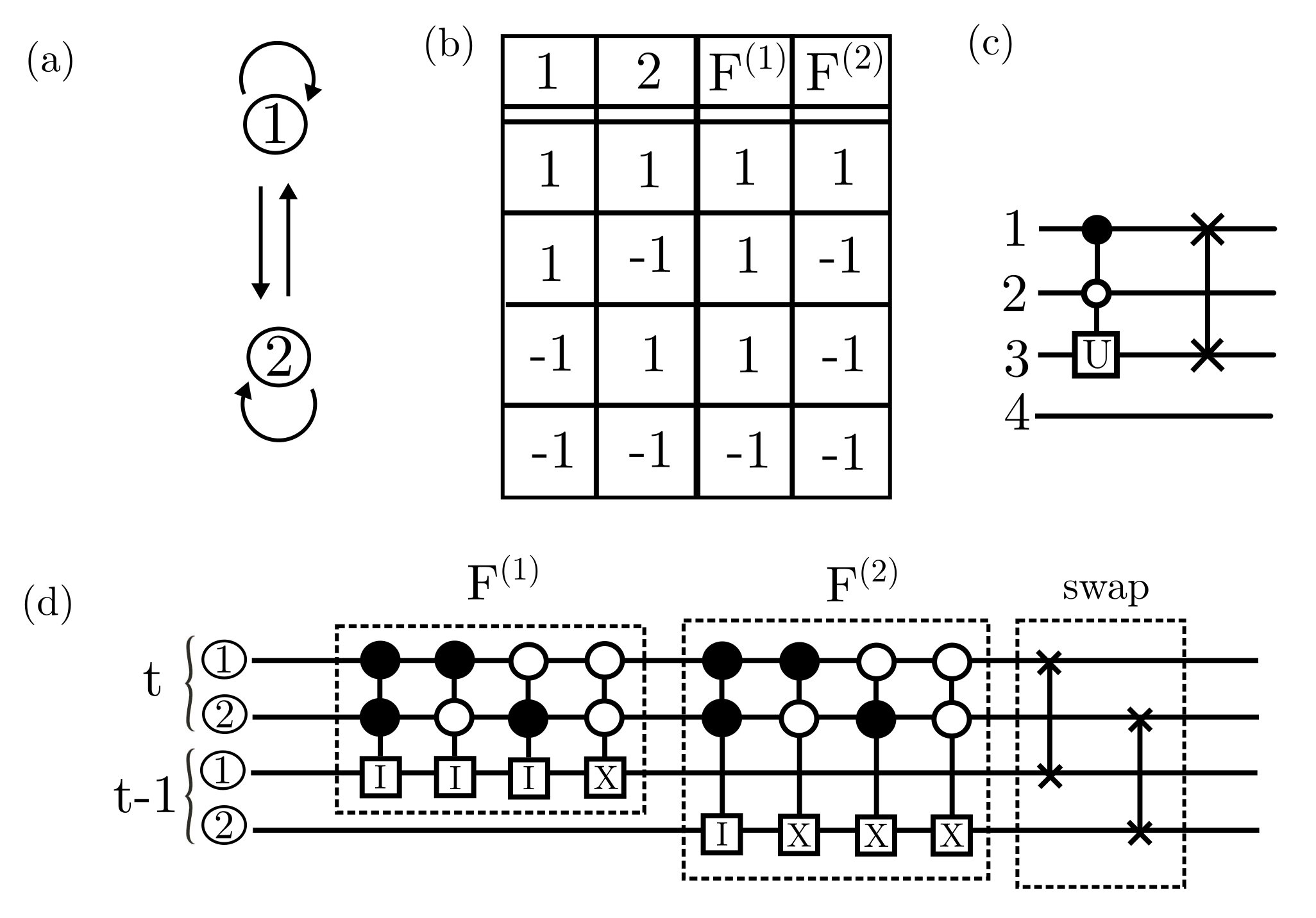}
\caption{Classical circuit of a two-node reversible Boolean network with a connectivity of two. Top shows (a) a classical representation of the network, (b) with related truth table for the binary logic functions $F^{(1)}$ (OR) and $F^{(2)}$ (AND). (c) fundamental logic gates, operation $U$ on the left and swap of two qubits on the right. The operation $U$ applied to qubit 3 is controlled by the qubits 1 and 2. (d) corresponding logic circuit. The upper two qubits represent the state at time $t$, while the lower ones correspond to the time $t-1$. I is the Identity operator, and X is the Pauli-X (spin-flip) operator. 
Operations are executed sequentially from left to right to execute the update rule (\ref{eq:reversible}). 
}
\label{fig:network}
\end{figure}

A simple example circuit is the $N=2$, $K=2$ case shown in Fig.~\ref{fig:network}.  It consists of two nodes: $s^{(1)}$ is updated according to the OR function $s^{(1)} \lor s^{(2)}$; and $s^{(2)}$ is updated according to the AND function $s^{(1)} \land s^{(2)}$.  Fig.~\ref{fig:network}(a) shows the wiring diagram indicating which nodes are inputs to each node, and (b) shows the results of applying Boolean functions $F^{(1)}$ and $F^{(2)}$ to each possible configuration of input states, i.e. the truth table.

Following Coppersmith et al., we choose to study the case in which the Boolean variables are updated synchronously.  To perform the necessary operations, we first introduce one auxiliary bit for each node in the network.  This bit holds the updated value of the node until all the operations at a given time step are completed, after which the values of the original bit and the corresponding auxiliary bit are swapped so that the original qubit assumes a new value and the auxiliary qubit holds the value of the original on the previous time step, and the operations for the next time step can begin.

An elementary operation is represented by the diagram in Fig.~\ref{fig:network}(c), where each line represents one bit, with lines 3 and 4 representing the auxiliary bits corresponding to bits 1 and 2, respectively. (For simplicity, we have relabeled $s^{(1)}_t$, $s^{(2)}_t$ as $s^{(1)}$, $s^{(2)}$ and $s^{(1)}_{t-1}$, $s^{(2)}_{t-1}$ as $s^{(3)}$, $s^{(4)}$.) 
Time proceeds from left to right, and the open and solid dots on lines 1 and 2 indicate that the operation $U$ is {\em controlled} by the values of those bits. Open and solid dots correspond to states $s = -1$ or $+1$, respectively. In order for $U$ to be applied to $s^{(3)}$,  $s^{(1)}$ and $s^{(2)}$ must take the specified values at time $t$.  In this example, $s^{(3)}$ just after time $t$ will be the result of applying $U$ if and only if $s^{(1)} = 1$ and $s^{(2)} = -1$.  Otherwise, $s^{(3)}$ will remain unchanged.
The $\times$ symbols connected by a vertical line indicate the bit values are swapped.

A Boolean function is represented as a sequence of $2^K = 4$ operations, each implementing a single row in the corresponding truth table.  We use the symbol $I$ to represent an identity operation and $X$ to represent a bit-flip (spin-flip).  The Boolean logic is represented as shown in Fig.~\ref{fig:network}(d), where the four operations in the box labeled $F^{(1)}$ leave the auxiliary bit $s^{(3)}$
unchanged or flip this state, depending upon the  value of $s^{(1)}\lor s^{(2)}$.  
The analogous procedure is applied to $s^{(4)}$ in the box labeled $F^{(2)}$, depending upon the value of $s^{(1)}\land s^{(2)}$. To complete a single time step, two swap operations are performed, so that $s^{(1)}$  and $s^{(2)}$ (representing the values at time t) now take the updated values of $s^{(3)}$ and $s^{(4)}$, respectively, and $s^{(3)}$ and $s^{(4)}$ take the values of $s^{(1)}$ and $s^{(2)}$ on the previous time step.
%\lk{Increasing the time step by one s (1) and s (2) become values at time t-1 and s (3) and s (4) will become values at time t.}

The classical logic operations can be described as follows, using a notation that generalizes to quantum logic operations.  We take each node to be a two-level system (qubit) and work in the computational basis $\ket{0}$ and $\ket{1}$, where $s = -1$ and $+1$ correspond to $\ket{0}$ and $\ket{1}$, respectively. 
The state of an isolated qubit is denoted by $\xi$.
The identity operation, denoted $I$, leaves the state unchanged.  The bit-flip operation takes $\xi=\ket{0}$ to $\xi=\ket{1}$ and $\ket{1}$ to $\ket{0}$, which is the action induced by the Pauli-X operator, denoted in the diagram by $X$.  (In $2 \times 2$ matrix notation, $X$ is represented by the Pauli matrix $\sigma_x$.)  If the only operations used in the circuit are (controlled) $I$ and (controlled) $X$, each qubit is always in one of the two computational basis states, and the dynamics is equivalent to a classical, reversible Boolean network.

The swap operation also generalizes to quantum systems, as the values of two qubits can be exchanged without loss of information~\cite{Nielsen:2011:QCQ:1972505}. We use the symbol $W$ to denote a swap, 
where $W(\xi^{(1)}\otimes \xi^{(2)}) = \xi^{(2)}\otimes \xi^{(1)}$, e.g., 
$W \ket{01} = \ket{10}$, where $\ket{01}$ stands for $\ket{0}\otimes \ket{1}$.

To introduce nontrivial quantum superpositions and interference effects, we add one additional operation to our set: the Hadamard operation $H$ that takes $\ket{0}$ to $\tfrac{1}{\sqrt{2}}(\ket{0} + \ket{1})$ and $\ket{1}$ to $\tfrac{1}{\sqrt{2}}(\ket{0} - \ket{1})$, corresponding to the matrix operation $\tfrac{1}{\sqrt{2}}(\sigma_x + \sigma_z)$.  
We note that the Hadamard operation in combination with classical operations can generate entangled states of the network. For example, beginning with a state of two qubits $\ket{00}$, applying $H$ to the first qubit produces the state $\tfrac{1}{\sqrt{2}}(\ket{00}+\ket{10})$. If $X$ is then applied to the second qubit while controlled by the first, we obtain the entangled state $\tfrac{1}{\sqrt{2}}(\ket{00}+\ket{11}$).
As explained below, the restriction of quantum operations to $I$, $\sigma_x$, and $H$ allows us to introduce a measure analogous to the Hamming distance between two network states. We will introduce this measure in Section~\ref{sec:hamming}.

In the computational basis, it is clear that the controlled operations $F^{(i)}$, which each leave all $\xi_t^{(i)}$ unchanged and change only the auxiliary qubit at node $i$, can be performed in any order within a time step as long as the swap operations are all delayed until the end of the step.  Under this protocol, each time step corresponds to a synchronous update of all of the $\xi^{(i)}$.  For each $i$, the swap causes the auxiliary qubit at node $i$ at the end of step $t$ to takes the value $\xi_t^{(i)}$, preparing it to be operated upon in step $t+1$.

We use the symbol ${\cal U}$ to represent the unitary propagator that advances the entire system through one complete time step.  ${\cal U}$ acts on a vector consisting of $2^{2N}$ basis states for the system of $N$ nodes, each of which has two associated qubits.  Each basis state has the form $\xi^{(1)}\otimes\xi^{(2)}\otimes\ldots \xi^{(2N)}$, where $\xi^{(i+N)}$ is the state of the auxiliary qubit at node $i$.  For the classical operations, the set of all states accessible from a given initial state consists entirely of the finite set of basis states, which immediately implies that the trajectory of the system on repeated application of ${\cal U}$ must be periodic.

\begin{figure}
\centering
 \includegraphics[width=0.7\linewidth]{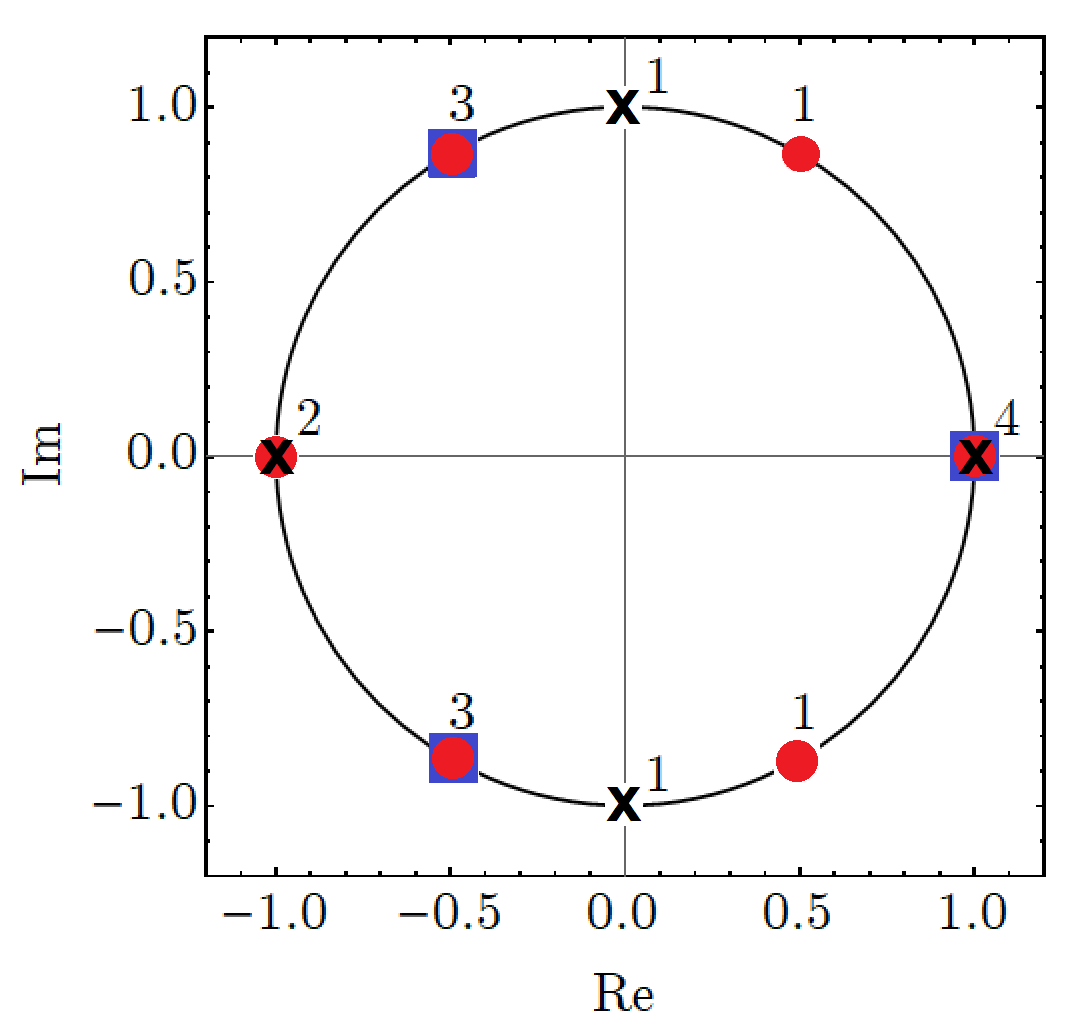}
  \caption{Complex eigenvalues of the classical circuit in Fig.~\ref{fig:network}. The eigenvalues corresponding to the cycle of length 6 are marked as red dots, those corresponding to the cycle of length 4 are marked as black crosses, and the two cycles of length 3 are marked as blue squares. Numbers indicate the degeneracies of the eigenvalues.}
 \label{fig:EVclassic}
\end{figure}

Reversibility ensures that each classically accessible state appears in exactly one periodic cycle, and a matrix representation of ${\cal U}$ for a classical network must simply be a permutation matrix with block diagonal form, where each block operates on the subspace of states that form a single cycle. 
The characteristic polynomial of a single block of dimension $L$ is:
\begin{align}
    p(\lambda) = &\begin{vmatrix}
    -\lambda & 1 & \cdots & 0 \\
\vdots & -\lambda &  1 & \vdots \\
\vdots &  &  \ddots& 1 \\
1 & \cdots & \cdots & -\lambda \\
    \end{vmatrix} \\
    = &(-1)^L(\lambda^L - 1)\,,
    \label{eq:ews}
\end{align}\\
which has roots
\begin{equation}
   \lambda_k= e^{2\pi i k/L }
\end{equation}
for integer $k$ ranging from $0$ to  $L-1$.
The eigenvalues of ${\cal U}$ thus consist of the union of complete sets of the $L^{th}$ roots of unity for a set of values of $L$ that sum to $2N$. Given the full set of eigenvalues, one can uniquely determine all of the periods by iteratively identifying the largest value of $L$ and removing one set of eigenvalues corresponding to it.
As an example, Fig.~\ref{fig:EVclassic} shows the eigenvalues of the classical reversible Boolean network of Fig.~\ref{fig:network}.  Recall that Fig.~\ref{fig:network}(d) shows 4 qubits, each represented by a horizontal line, that store the values of two qubits for two time steps. Hence there are $2^4=16$ basis states, and one finds one cycle of length 6, one cycle of length 4 and two cycles of length 3. They correspond to the matrix blocks of dimension $L=6$, $L=4$, and $L=3$ (twofold), respectively, associated with the eigenvalues marked by red dots, black crosses, and blue squares, respectively.

\begin{figure}
  \centering
  \includegraphics[width=0.7\linewidth]{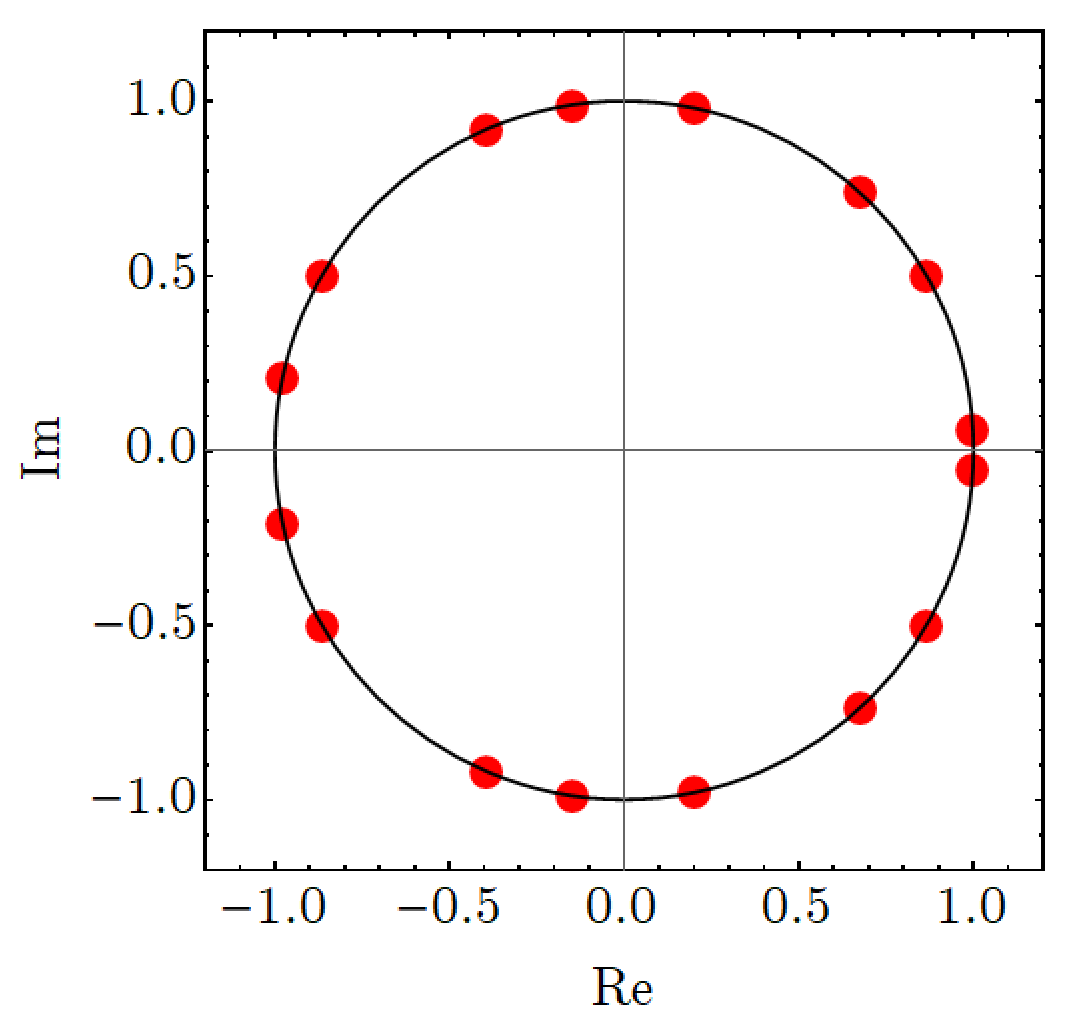}
    \caption{Complex eigenvalues of the circuit in Fig.~\ref{fig:network} with a Hadamard operation included. All eigenvalues are non-degenerate and have phases that are irrational multiples of $2\pi$.}
\label{fig:EVquantum}
\end{figure}

When the Hadamard operation is added to the set of logic operations, the distribution of eigenvalues of ${\cal U}$ takes a qualitatively different form; an example is shown in Fig.~\ref{fig:EVquantum}. The Hadamard operation is introduced before the first logic gate on qubit 1 at time t-1 (target-qubit) and is applied on every update step ${\cal U}$. First, while unitarity ensures that all eigenvalues fall on the unit circle, the eigenvalues are now generally not roots of unity. For networks with 2-input gates, the inclusion of the Hadamard operation makes an infinite set of superposition states accessible~\cite{nebe2001invariants}, which can accommodate quasiperiodic trajectories~\cite{aharonov2003simple} corresponding to eigenvalues with phases that are irrational multiples of $2\pi$. Second, there is no degeneracy in the spectrum in this example. 

The situation for $K=1$ networks is qualitatively different.  Here, the use of operations consisting only of $H$, $\sigma_x$, $I$, and $W$ ensures that ${\cal U}$ is a member of the Clifford group ${\cal C}_n$, defined as follows.  Let $r_{i,j,\ldots}$ be a tensor product of $2^N$ Pauli operators $\{ \sigma^{(1)}_i\otimes \sigma^{(2)}_j \otimes \ldots \}$. ${\cal C}_n$ consists of all unitary operators $U$ of dimension $2^N$ for which
\begin{equation}
 U r_{i,j,\ldots}U^{\dagger} = r_{i',j',\ldots}{\rm\ modulo\ } U(1)
\label{eqn:Clifford}
\end{equation}
for every $r_{i,j,\ldots}$, where $U(1)$ represents an overall phase factor of no physical significance.~\cite{gottesman1997stabilizer}. 
An important property of ${\cal C}_n$ is that it contains a finite number of elements \cite{ozols2008clifford, calderbank1998quantum}. As a result, one can show that ${\cal U}^m = {\cal U}$ for some integer $m$, which implies that all trajectories in these $K=1$ networks are periodic and the eigenvalues are roots of unity.
%\js{[NOTE: It would be nice to find a more straightforward reference for the Clifford group, given that the word ``Clifford'' does not appear anywhere in Gottesman's thesis. $\rightarrow$ \cite{calderbank1998quantum} Calderbank et al.]}

\section{Propagation of perturbations}

\subsection{Distances between states} \label{sec:hamming}
In this section we introduce a measure for characterizing how perturbations spread through our $K=1$ networks. For classical networks, the evolution of the Hamming distance is often used for this purpose.  It is defined as the number of nodes whose values differ between states at the same time step on two different trajectories.  Coppersmith et al.~\cite{coppersmith2001reversibleII} studied evolution of the Hamming distance~\cite{Wegner:1960:TCO:367236.367286} in reversible networks for two states that initially differ at only a single node.

A straightforward generalization of the Hamming distance to quantum states is useful when a state $\psi'$ can be written as $r'_{i,j\ldots} \psi$.
We define the distance between $\psi'$ and $\psi$ as the number of factors in the first $N$ terms in $\rho_{i,j\ldots}$ that differ from the identity. This definition reduces to the classical Hamming distance in cases where the elements in $r'_{i,j\ldots}$ include only the identity and the classical bit flips represented by $\sigma_x$.
The restriction to the first $N$ terms picks out the factors representing the current states of the qubits, leaving out the auxiliary qubits that represent the state at time $t-1$.

Given two states related at time $t=0$ by $\psi'_0 = r'_{i,j\ldots} \psi_0$  
and a set of allowed quantum operations that is restricted to elements of the Clifford group,
we have at time $t=1$:
\begin{align}
    \psi'_{1} & =  \mathcal{U}r'_{i,j\ldots}\psi_{0} \\
    \ & =  \mathcal{U}r'_{i,j\ldots}\mathcal{U}^\dagger\mathcal{U} \psi_{0} \\
    \ & =  \left(\mathcal{U}r'_{i,j\ldots}\mathcal{U}^\dagger\right) \psi_{1} \\
    \ & =  \rho_{i,j\ldots} \psi_{1}
    \label{eq:comp}
\end{align}
for some $\rho_{i,j\ldots}$, by Eq.~(\ref{eqn:Clifford}).
Thus the generalized Hamming distance remains a useful measure of the distance between the future trajectories of the states.

We emphasize again that this extended version of the Hamming distance applies only for networks in which each gate has at most a single input. We also note that any entanglement arising in one trajectory is necessarily mirrored in the other, as the elements of $\rho_{i,j\ldots}$ act on single qubits, preserving any entanglement among the different qubits.

\subsection{Components of $K=1$ networks}

A network drawn from a random ensemble with fixed $K$ may contain subsets of nodes forming connected components that have no effective inputs from any other nodes and no outputs to any other nodes.  An input to node $i$ from node $j$ is {\em not} effective if the value of $F^{(i)}$ is completely independent of $\xi_{t}^{(j)}$.

The numbers and sizes of these independent components strongly depend on the in- and out-degree distributions of the network.
The dynamics generated on a given connected component is completely independent of the dynamics of the rest of the network, and it is clear that a perturbation occurring at a single node in a random Boolean network can propagate through only a single connected component.  We are therefore interested in the dynamics supported by a single connected component of a network; we leave aside the question of combinatorial effects arising from multiple independent perturbations applied in independent components.
 
\begin{figure}[t]
  \includegraphics[width=.7\linewidth]{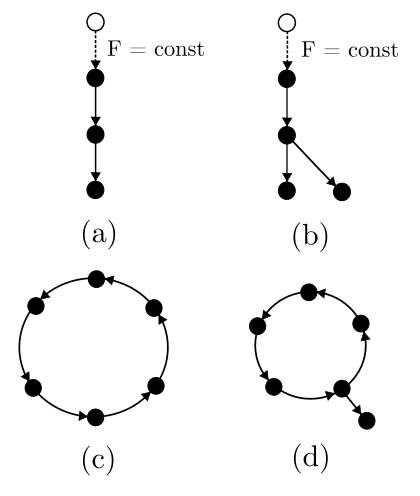}
  \caption{Examples of single network components. Black dots indicate nodes and arrows indicate directed inputs. (a) shows a chain, and (b) shows a chain with an additional node attached. For both cases, the chain starts with the node following the dotted arrow. The label "F = const" indicates that the first node of the chain (empty circle) is not affected at all by the value of the node supplying its input due to a trivial choice of the associated logic function F. (c) displays a loop and (d) a loop with an additional node attached.}
 \label{fig:net}
\end{figure}

We discuss here only components of $K=1$ networks, where each node has in-degree equal to one, which allows for an analysis of the spread of perturbations using the generalized Hamming distance measure.
For dissipative (classical) $K=1$ networks, a connected component must consist of a single loop of nodes with dead-end chains emanating from it (where the loop may be just a single node with a self-input).  The dynamics generated by such structures has been characterized in detail~\cite{flyvbjerg1988exact}.
Here, however, we are interested in reversible networks, which exhibit qualitatively different dynamics from their dissipative counterparts.
It is convenient to identify two special cases:
(1) a {\em chain}, in which the first node receives a constant input and the last has no output (see Fig.~\ref{fig:net}(a)), and (2) a {\em loop}, in which each node receives its input from a neighbor such that the set of input edges forms a closed ring (see Fig.~\ref{fig:net}(c)).~\cite{coppersmith2001reversible}
All more complex components consist of a single loop or a constant node with directed trees emanating from it. The simplest examples are shown in Fig.~\ref{fig:net}(b) and (d): a chain and a loop with one additional node added.

For reversible networks, the meaning of the specification ``$K=1$'' is slightly different because all nodes receive time-delayed self-inputs in addition to inputs from other nodes.  $K=1$ here means that each node has exactly one input coming from the value of a node at the current time. Using the implementation in which auxiliary nodes hold the information about time-delayed values, each node effectively receives $K+1$ inputs, one being the auxiliary bit. 

We emphasize here that the reversibility of the network dynamics is a logical property, not a guarantee of time-reversal symmetry.  The directed links in the network generally break time-reversal symmetry by introducing different mechanisms for propagating information forwards or backwards across any given link.

\begin{figure}[t]
  \includegraphics[width=\linewidth]{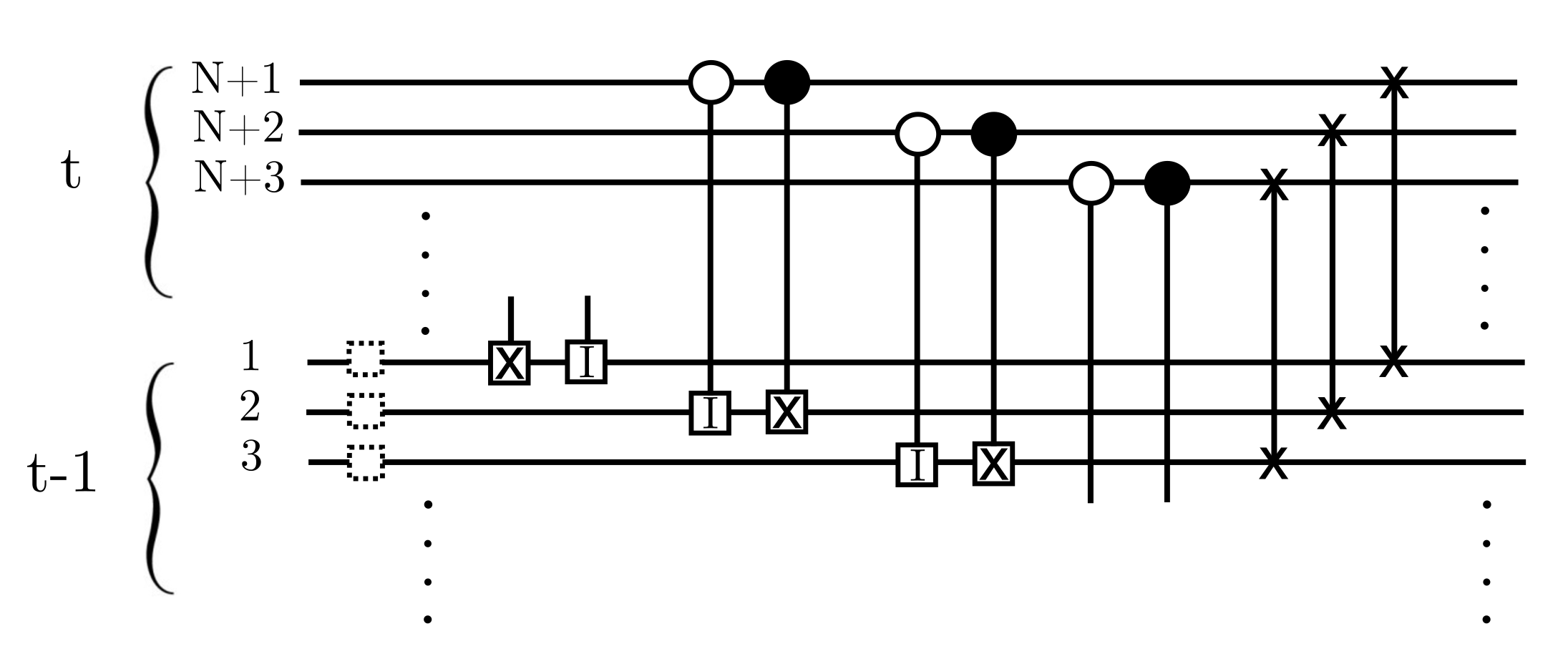}
  \caption{Logic circuit of a loop, consisting of N nodes. For simplicity, we only display the three first nodes, at time $t-1$ and $t$. Labels N+1, N+2, N+3 on the left of the qubit line denote the control bits at time $t$, whereas labels 1, 2, 3 denote the target bits at time $t-1$. The dotted squares denote the position at which the Hadamard operation will be inserted later on. Each pair of controlled operations, applied to the same bit, represents a truth function $F$. Each truth function $F^{(i)}$ uses bit $i-1$ as an input and bit $N+i$ as a target. For a loop the last control bit $N$ will work on the first target bit $1$.}
 \label{fig:loopScheme}
\end{figure}

To avoid confusion, we use the term ``loop'' for a set of nodes connected topologically in a circle, ``cycle'' for the time-periodic dynamics of a network in state space, and ``circuit'' to refer to a quantum network.

\subsection{Analysis of Spreading Perturbations}

\begin{figure*}[t]
  \includegraphics[width=\linewidth]{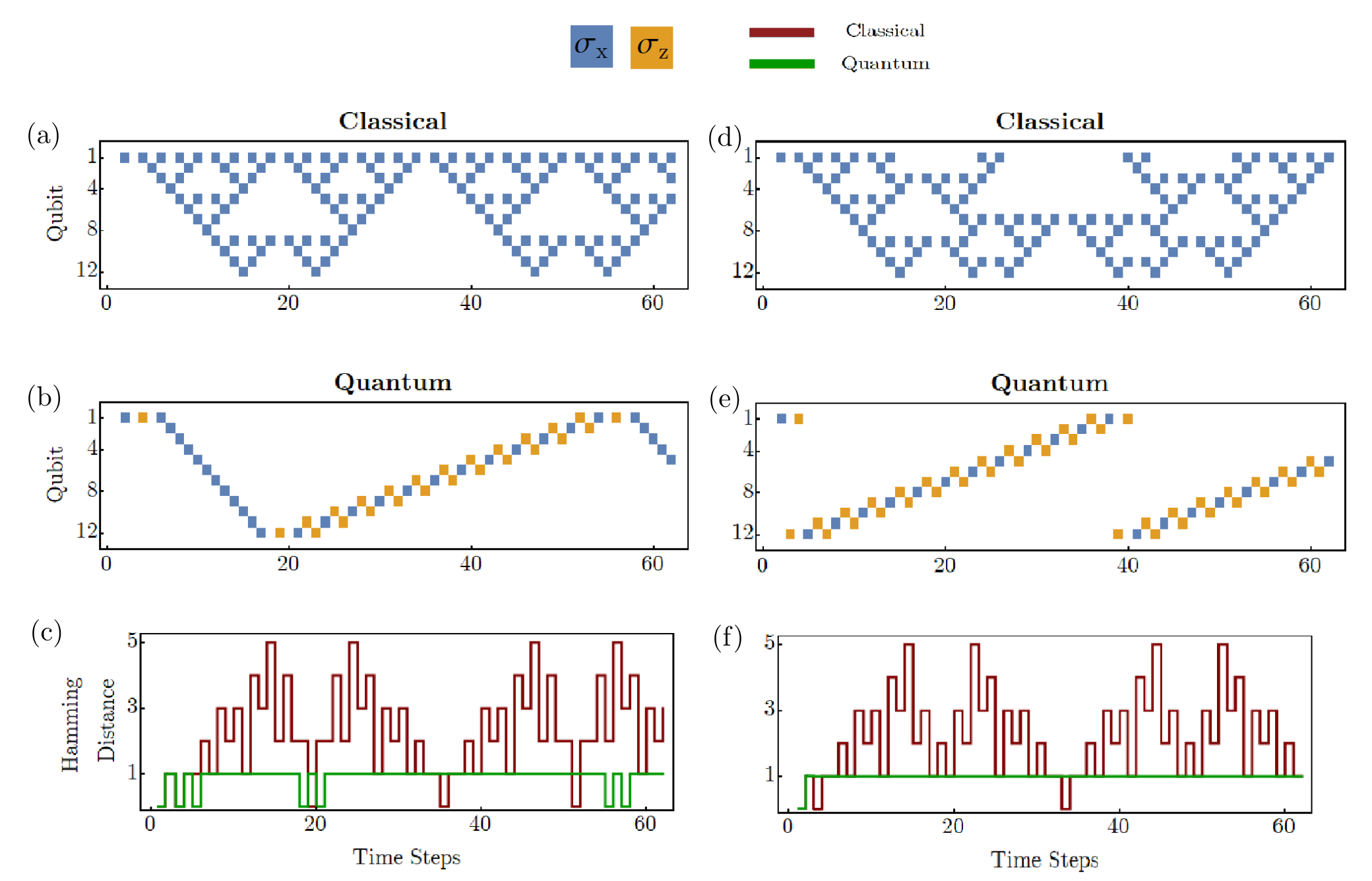}
  \caption{Spatial perturbation patterns and Hamming distance of a 12-qubit chain (a)-(c) and loop (d)-(f) plotted over time. The operators $\sigma_x, \sigma_y$ are marked in blue and orange, respectively. The panels labeled Classical, (a) and (d), show the spatial evolution in the classical network. A colored square corresponds to a Pauli operator applied to the related qubit. The panels labeled Quantum, (b) and (e), correspond to the quantum version. For the quantum systems one Hadamard operation has been applied to each of the target qubits before applying the logic function (compare Fig.~\ref{fig:network}). Panels (c) and (f) display the total Hamming distance of both systems (red: classical, green: quantum), which can be obtained by summing over all squares in one time step. The Hamming distance is always measured after the application of the final Swap operations.}
  \label{fig:propagation}
\end{figure*}

In this subsection we analyze the spatial propagation of small perturbations within a single network component.
Table~\ref{tab:Clifford} shows the transformation of the Pauli matrices under the Hadamard operation.  Because our logic operations $F^{(i)}$ involve only $\sigma_x$ and $I$, and because the perturbations considered are simple bit flips induced by $\sigma_x$, the $r_{i,j,\ldots}$ operation relating the perturbed and original trajectories contains only $\sigma_x$ and $\sigma_z$ terms.
Fig.~\ref{fig:propagation} shows the classical and quantum mechanical spatial perturbation pattern for a chain (left column) and a loop (right column). Colored squares indicate differences between two trajectories that initially differ by a single bit flip. Blue and orange indicate $\sigma_x$ and $\sigma_z$ factors in $r_{i,j,\ldots}$. 
\begin{table}[t]
    \centering
 \begin{tabular}{c || c c c }
 $\sigma_i$& $\sigma_x$ & $\sigma_y $& $\sigma_z$ \\
 \hline \hline
  $H\sigma_i H^{\dagger}$ &$\sigma_z$&$-\sigma_y$&$\sigma_x$
 \end{tabular}
 \vspace{12pt}
\caption{Mapping rules of the Hadamard gate (H) for Pauli matrices $\sigma_i$.}
\label{tab:Clifford}
\end{table}

Coppersmith et al.\cite{coppersmith2001reversibleII} already studied the pattern generated by small perturbations in the classical networks. As displayed in Fig.~\ref{fig:propagation}(a),(d), the pattern of the classical Hamming distance shows $90^{\circ}$-rotated Sierpinski gaskets \cite{mandelbrot1977fractals}, which are generated by rule 90 of the automata scheme introduced in \cite{wolfram}. Using our notation, the Boolean version of rule 90 can be written as
\begin{equation}
s^{(i+1)}_{t} =  s^{(i)}_{t+1} s^{(i)}_{t-1}.
\label{eq:rule90}
\end{equation}
For our observed pattern, the time and space axes are exchanged with respect to the typical depiction of the rule 90 pattern. 
To show that this pattern results from Eq.\eqref{eq:reversible}, Coppersmith et al.\ consider two trajectories, ${\bf s}_t$ and ${\bf s'}_t$, on a chain or loop, where the trajectories begin from initial configurations that differ in one bit; $s^{(1)}_0 s'^{(1)}_0 = -1$.
Next, they define the product of the two solutions as ${\bf r}_t$, with $r^{(i)}_t = s^{(i)}_t s'^{(i)}_t$.
Rewriting Eq.~\eqref{eq:reversible} for the case of a chain or loop we obtain
\begin{equation}
        s_{t+1}^{(i)} =  s_{t-1}^{(i)}F^{(i)}\left(s^{(i-1)} _t\right),
        \label{eq:Fchain}
\end{equation}
where $F^{(i)}(s^{(i-1)} _t)=\pm s^{(i-1)}_t$.
The evolution of perturbations $r^{(j)}_t$ can then be expressed as 
\begin{equation}
    r^{(j)}_{t+1} = r^{(j)}_{t-1} r^{(j-1)}_t,
    \label{eq:perturb}
\end{equation}
which mirrors Eq.\eqref{eq:rule90}.
Note that the evolution of the perturbation is independent of the initial values of the bits and also independent of the functions $F^{(i)}$.

The evolution of a perturbation in the quantum network can be written as follows using the notation defined above.
We denote $\rho_{i,j\ldots}$ in Eq.~\eqref{eq:comp} by $\{ \rho^{(i)}_1\otimes \rho^{(j)}_1 \otimes \ldots \}$, or, after multiple time steps: 
\begin{equation}
\rho_{i,j\ldots,t} = \rho^{(i)}_t\otimes \rho^{(j)}_t \otimes \ldots .
\end{equation}
Let $C$ and $\tilde{C}$ represent the controlled-not gates activated by the control qubit being in state $\ket{1}$ or $\ket{0}$, respectively.
The propagator ${\cal U}$ consists of a sequence of operations $u_k$ that may include $I$ or $H$ acting on single qubits, $C$ or $\tilde{C}$ acting on pairs of qubits, and finally a set of $W$ (swap) operations.
Consider the action of one of these operations on each of the trajectories we are comparing.  
A given $I$, $H$, or $W$ operation clearly has no effect on $\rho_{i,j\ldots,t}$, as it produces the same effect on both trajectories.
One can also confirm by direct enumeration that the effects of $C$ and $\tilde{C}$ on $\rho_{i,j\ldots,t}$ are the same up to an overall sign that is physically irrelevant; i.e.,
\begin{align}
    C\cdot [\rho^{(i)}\otimes \rho^{(j)}]\cdot C^{\dagger} = \pm \tilde{C} \cdot [\rho^{(i)}\otimes \rho^{(j)}]\cdot \tilde{C}^{\dagger}
\end{align}
for any $\rho^{(i)}$ and $\rho^{(j)}$ in the set $\{I, \sigma_x, \sigma_y, \sigma_z\}$.
Thus, as long as the $W$ operations are all performed after all of the others, we see that $\rho_{i,j\ldots,t}$ for our $K=1$ networks does not depend on the logic functions associated with the directed links in the network graph.  
(Note that this is qualitatively similar to that of {\em dissipative} $K=1$ loops, where a single bit-flip creates a difference between two trajectories that simply propagates around the loop one step at a time, and the difference between the two is independent of whether a link performs a Copy or a Not operation.)

We now investigate the changes induced when quantum operations are added, 
%In both cases, classical and quantum, the initial
considering perturbations applied to an initial state $\psi_{t=0}$ (compare Eq.~\eqref{eq:reversible}). From the 2N qubits needed to calculate the evolution of the network, Fig.~\ref{fig:propagation}(b),(e) displays qubits 1,\dots, N for every time step after the application of the Swap operations. In the quantum case, Hadamard operations are applied to each target qubit before applying the logic function. 
Both types of component, the chain and the loop, exhibit remarkable spatial propagation patterns that are substantially less complex than their classical counterparts.
In the following we will discuss the quantum patterns in more detail. They are obtained by calculating Eq.~\eqref{eq:perturb} for each operation within the network. 

For present purposes, we limit our investigation to networks in which the Hadamard operator $H$ is applied at every target qubit once per time step. This subset contains networks that show markedly different properties than the networks realizable by classical operations alone.  We refer to the networks with Hadamard operators as ``quantum networks'' and to those with no Hadamard operators as ``classical networks.''  Exploration of the behavior of mixed cases in which $H$ is applied only to a subset of the qubits is beyond the scope of the present paper.

We begin by studying the evolution of a perturbation on a simple loop.  Surprisingly, it spreads in only one direction, as shown in  Fig.~\ref{fig:propagation}(e).  The following pattern emerges: \\

\begin{align}
     \sigma^{(1)}_x  &\xrightarrow{H}   \sigma^{(1)}_z \xrightarrow{F}  \sigma^{(1)}_z \otimes \sigma^{(2N)}_z  \xrightarrow{W}  \sigma^{(N)}_z \otimes \sigma^{(N+1)}_z\label{eq:circ11} \\
     & \xrightarrow{H}  \sigma^{(N)}_x \otimes \sigma^{(N+1)}_z 
      \xrightarrow{F,W}  \sigma^{(1)}_z \otimes \sigma^{(2N)}_x \\
     & \xrightarrow{H}  \sigma^{(1)}_x \otimes \sigma^{(2N)}_x 
     \xrightarrow{F} \sigma^{(2N)}_x 
     \xrightarrow{W}  \sigma^{(N)}_x  \label{eq:circ112}\\
     &\xrightarrow{H}  \sigma^{(N)}_z
     \xrightarrow{F}  \sigma^{(N)}_z \otimes \sigma^{(2N-1)}_z \nonumber \\ 
     &\xrightarrow{W}  \sigma^{(N-1)}_z \otimes \sigma^{(2N)}_z \\
     &\xrightarrow{H}  \sigma^{(N-1)}_x \otimes \sigma^{(2N)}_z   \xrightarrow{F,W}  \sigma^{(N)}_z \otimes \sigma^{(2N-1)}_x \\ 
    &\xrightarrow{H}  \sigma^{(N)}_x \otimes \sigma^{(2N-1)}_x   \xrightarrow{F}   \sigma^{(2N-1)}_x \xrightarrow{W}   \sigma^{(N-1)}_x .\label{eq:circ12}
\end{align}

Each $\sigma_i$ denotes a Pauli matrix, representing a term in the tensor product $\rho_{i,j\ldots}$ that relates the trajectory states $\psi_{t}$ and $\psi'_{t}$.  All terms that are not explicitly listed are identity elements. 
The operations applied to the qubits are denoted by letters on top of the arrows. The presence of two letters on the same arrow indicates that the first operation does not change the state. Superscripts $1,\dots,N$ denote the target qubit, and $N+1,\dots,2N$ denote the control qubits. The Pauli operators transform under the application of a single-input gate as indicated in Tab.~\ref{tab:CNOT}. Note that $Z$ elements are produced by Hadamard operations on single qubits and that $Y$ elements are produced when the control bits differ by $X$ and the targets by $Z$.

\begin{table}[t]
    \centering
 \begin{tabular}{c ||@{\hspace{1em}}c@{\hspace{1em}} |@{\hspace{1em}}c@{\hspace{1em}} |@{\hspace{1em}}c@{\hspace{1em}} |@{\hspace{1em}} c@{\hspace{1em}} }
  \multirow{2}{*}{Control} & \multicolumn{4}{c}{Target} \\
    & $I$ & $X$ & $Y$ & $Z$ \\
 \hline \hline
\rule[-5pt]{0pt}{15pt}$I$ & $I \otimes I$ & $I \otimes X$ &$Z \otimes Y$ & $Z \otimes Z$ \\ \hline
\rule[-5pt]{0pt}{15pt}$X$ & $X \otimes X$ & $X \otimes I$ & $Y \otimes Z$ & $-Y \otimes Y$  \\ \hline
\rule[-5pt]{0pt}{15pt}$Y$ & $Y \otimes X$ & $Y \otimes I$ & $-X \otimes Z$ & $X \otimes Y$ \\ \hline
\rule[-5pt]{0pt}{15pt}$Z$ & $Z \otimes I$ & $Z \otimes X$ & $I \otimes Y$ & $I \otimes Z$  \\ 

\end{tabular}
\vspace{10pt}
\caption{Mapping rules under the action of a controlled-X gate for tensored Pauli operators representing differences between two trajectories.
Each process can be reversed by another application of the controlled-X gate. Row labels represent control qubits; column labels refer to the target qubit. Table entries are given in the form ``control $\otimes$ target.'' $X$, $Y$, and $Z$ represent $\sigma_x$, $\sigma_y$, and $\sigma_z$.}
\label{tab:CNOT}
\end{table}

Each line in Eqs.\eqref{eq:circ11}-\eqref{eq:circ12} corresponds to a single time step, which is concluded by the  $W$ (swap) operation. 
For illustration we discuss in detail the first time step. The Pauli-X operator $\sigma_x^{(1)}$, describing the difference between the first two target qubits, i.e., at times $t=0$ and $t=1$, transforms upon the application of a Hadamard operation into a $\sigma_z$ operator which stays in the same position $\sigma_z^{(1)}$. Next, we apply a non-trivial truth function $F$ which is input-dependent. Both possible non-trivial truth functions $C$ and $\tilde{C}$
yield the same outcome, so $F$ need not be further specified. We can use Tab.~\ref{tab:CNOT} to show that, since $\sigma_z$ is a perturbation on a target qubit, the application of a truth function generates yet another $\sigma_z$ operator on the $2N$-th qubit ($\sigma_z^{(2N)}$), which is a control qubit. Finally, all target and control qubits are swapped (W).
%It can be seen that it takes the network three steps to reach the following bit {\bf ???}. 
Considering Fig.~\ref{fig:propagation}(e), we observe that one period of the 12-node loop takes 36 time steps. 

We find that, contrary to the classical networks where the perturbations spreads over the whole network component, the quantum system exhibits a localized perturbation which moves through the system affecting only one qubit at a time. We will refer to the configuration of Pauli operators that propagates in this way as a {\em solitary state}. Equations.~\eqref{eq:circ11}-\eqref{eq:circ12} show how the solitary state moves through the system. The general, 3-time-step, form can be expressed as $\sigma^{(i)}_x \rightarrow \sigma^{(i-1)}_z \otimes \sigma^{(N + i)}_z \rightarrow \sigma^{(i)}_z \otimes \sigma^{(N+i-1)}_x \rightarrow \sigma^{(i-1)}_x$, where each arrow represents a full time step.

This propagation of a highly localized perturbation in the reversible quantum loop is reminiscent of the situation in classical {\em dissipative} $K=1$ networks, where the only structures supporting nontrivial dynamics are loops of COPY and INVERT gates. In the reversible classical loops, a perturbation simply propagates to the next node in the loop on every time step, as the value of a given node is completely determined by the value of its input node on the previous time step.  The mechanism sustaining a solitary state in the reversible quantum network, however, relies on quantum coherences between the primary and auxiliary qubits.

Next, we investigate the chain, displayed in Fig.~\ref{fig:propagation}(a)-(c). Here a solitary state first propagates in the opposite direction and with a different velocity than the one observed on the loop.  Instead of requiring three time steps, the new pattern moves to the next qubit in just one time step.
Equations~\eqref{eq:link}-\eqref{eq:link12} display the detailed propagation pattern of a chain with an initial perturbation applied at node 1. Each time step is concluded by the application of a swap operation and is indicated with an equation label. In total we display six time steps.
\begin{align}
\sigma_x^{(1)} &\xrightarrow{\text{H}} \sigma_z^{(1)}  \xrightarrow{\text{F}} \sigma_z^{(1)}  \xrightarrow{\text{W}} \sigma_z^{(N+1)} \label{eq:link} \\
&\xrightarrow{\text{H}} \sigma_z^{(N+1)}  \xrightarrow{\text{F}} \sigma_z^{(N+1)}  \xrightarrow{\text{W}} \sigma_z^{(1)}\\
&\xrightarrow{\text{H}} \sigma_x^{(1)} \xrightarrow{\text{F}} \sigma_x^{(1)}  \xrightarrow{\text{W}} \sigma_x^{(N+1)}\\
&\xrightarrow{\text{H}} \sigma_x^{(N+1)}  \xrightarrow{\text{F}} \sigma_x^{(2)} \otimes \sigma_x^{(N+1)} \nonumber \\  
&\xrightarrow{\text{W}} \sigma_x^{(1)} \otimes \sigma_x^{(N+2)}\\
&\xrightarrow{\text{H}} \sigma_z^{(1)} \otimes \sigma_x^{(N+2)} \xrightarrow{\text{F}} \sigma_z^{(1)} \otimes \sigma_x^{(3)}\otimes \sigma_x^{(N+2)} \nonumber  \\ 
&\xrightarrow{\text{W}} \sigma_x^{(2)} \otimes \sigma_z^{(N+1)} \otimes \sigma_x^{(N+3)}\label{eq:link11}  \\
&\xrightarrow{\text{H}} \sigma_z^{(2)} \otimes \sigma_z^{(N+1)} \otimes \sigma_x^{(N+3)} \nonumber\\
&\xrightarrow{\text{F}} \sigma_z^{(2)} \otimes \sigma_x^{(4)}\otimes \sigma_x^{(N+3)} \nonumber \\
&\xrightarrow{\text{W}} \sigma_x^{(3)} \otimes \sigma_z^{(N+2)} \otimes \sigma_x^{(N+4)}  \label{eq:link12}
\end{align}
A different solitary state emerges because the first qubit does not have an input. 
It takes the system five time steps to reach the solitary state, which propagates opposite to the direction of the solitary state produced on the loop. A full propagation step of this state is given in Eq.~\eqref{eq:link12}. We see that the Pauli-X operator propagates one bit with every time step. In general this solitary state can be written as: $\sigma_x^{(i)} \otimes \sigma_z^{(N+i-1)} \otimes \sigma_x^{(N+i+1)}$ for $i>1$ and with every time step $i$ is increased by one.

When this solitary state reaches the end of the chain, 
%we observe the same problem as in the beginning. The last qubit does not give input to any other qubits and the solitary state cannot propagate any further. 
it takes the system a few time steps to reach the same solitary state as was seen on the loop, described by Eqs.~\eqref{eq:circ112}-\eqref{eq:circ12}. The perturbation then reflects again off of the end of the chain, forming a periodic cycle. In total, one period takes approximately $4N$ time steps (neglecting the transition process between two solitary states).

We note that the Hamming distance for both the chain in Fig.~\ref{fig:propagation}(c) and the loop in Fig.~\ref{fig:propagation}(f) never exceeds unity in the quantum case, whereas in the classical case the Hamming distance oscillates between zero and five for the $N=12$ case.  In general, the maximum Hamming distance is a complicated function of $N$.  For $N=2^{n-1}$, however, we have observed that it is simply the $n^{th}$ Fibonacci number. 

\begin{center}
\begin{figure*}
  \includegraphics[width=.85\linewidth]{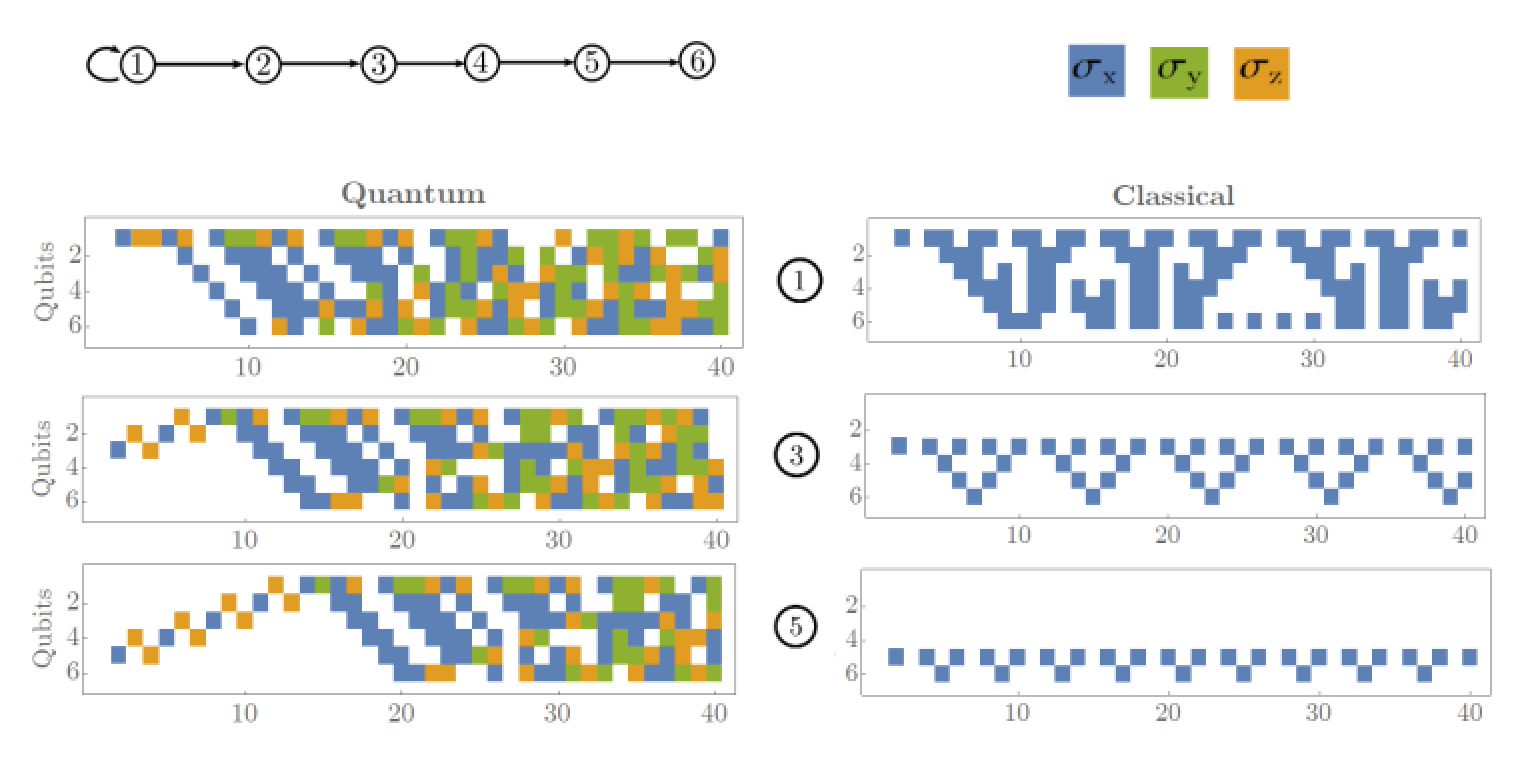}
  \caption{Time evolution of small perturbation for different initial conditions within the structure shown as inset on the top left. The left column displays the quantum system with Hadamard operations applied to all qubits, whereas the right columns shows the classical system. For each row the initial perturbation was introduced into a different qubit. In the first row the qubit 1 is initially perturbed, in the second row the qubit 3 and in the third row the qubit 5 are initially perturbed. The Hamming distance is always measured after the application of the final Swap operations.}
  \label{fig:ics}
\end{figure*}
\end{center}

Next, we briefly discuss the perturbation pattern of a  slightly more complex network component. We consider the 5-node chain attached to a 1-node loop, displayed in Fig.~\ref{fig:ics}. Two findings should be pointed out in particular. First, if the network component deviates from a plain loop or chain, the perturbation pattern becomes increasingly complex and can strongly differ from the ones shown for loops and chains. We will come back to that point in Section~\ref{sec:hammingstats}.

Second, the nature of the perturbation changes qualitatively as $\sigma_y$ elements appear in $\rho_{i,j\ldots}$. If a controlled $\sigma_x$ gate is applied to the combination of $\sigma_x$ (control) and $\sigma_z$ (target) perturbation, the result is  $\sigma_y$ matrices on both the control and target qubit (Tab.~\ref{tab:CNOT}). In the following we display the equations describing the time steps of the propagation of a perturbation on this network, showing how a Pauli-Y matrix can arise. We consider the case where the perturbation is introduced on the third target qubit (Fig.~\ref{fig:ics} middle row):
\begin{align}
     \sigma^{(3)}_x  &\xrightarrow{\text{H}}   \sigma^{(3)}_z \xrightarrow{\text{F}}  \sigma^{(3)}_z \otimes \sigma^{(N+2)}_z
     \xrightarrow{\text{W}}  \sigma^{(2)}_z \otimes \sigma^{(N+3)}_z\\
     &\xrightarrow{\text{H}}   \sigma^{(2)}_x  \otimes \sigma^{(N+3)}_z \xrightarrow{\text{F,W}} \sigma^{(3)}_z \otimes \sigma^{(N+2)}_x\\    
    &\xrightarrow{\text{H}}   \sigma^{(3)}_x \xrightarrow{\text{F}}  \sigma^{(N+2)}_x \xrightarrow{\text{W}} \sigma^{(2)}_x\\
    &\xrightarrow{\text{H}}   \sigma^{(2)}_z \xrightarrow{\text{F}}  \sigma^{(2)}_z \otimes \sigma^{(N+1)}_z
     \xrightarrow{\text{W}}  \sigma^{(1)}_z \otimes \sigma^{(N+2)}_z\\
     &\xrightarrow{\text{H}}   \sigma^{(1)}_x \otimes \sigma^{(N+2)}_z \xrightarrow{\text{F,W}} \sigma^{(2)}_z \otimes \sigma^{(N+1)}_x\\
    &\xrightarrow{\text{H}}   \sigma^{(2)}_x \otimes \sigma^{(N+1)}_x \xrightarrow{\text{F}}   \sigma^{(1)}_x \otimes \sigma^{(N+1)}_x \nonumber \\ 
    &\xrightarrow{\text{W}}   \sigma^{(1)}_x \otimes \sigma^{(N+1)}_x\\
    &\xrightarrow{\text{H}}   \sigma^{(1)}_z \otimes \sigma^{(N+1)}_x \xrightarrow{\text{F}}   \sigma^{(1)}_y \otimes \sigma^{(2)}_x \otimes 
    \sigma^{(N+1)}_y \nonumber \\ 
    &\xrightarrow{\text{W}}   \sigma^{(1)}_y \otimes \sigma^{(N+1)}_y \otimes \sigma^{(N+2)}_x.
\end{align}

Recall that all figures of perturbation patterns show only the target qubits. Control qubits are not shown, but they are needed to determine the network evolution. For all cases shown, our initial perturbation is a single $\sigma_x$, with no further perturbation on neither target nor control bits. Adding ``invisible'' perturbations on the control bits can change the behaviour significantly. For example, considering a simple loop and initializing it with $\sigma_x^{(2)} \otimes \sigma_z^{(2N)} \otimes \sigma_x^{(N+2)}$ perturbation instead of simply $\sigma_x^{(1)}$ would result in the solitary state described in Eq.\eqref{eq:link11}-\eqref{eq:link12}, even though the initial perturbation would look the same in our plots.

Considering the fact that in the classical system only $\sigma_x$ matrices are allowed as perturbations, there is only a single mechanism allowing perturbations to spread. If a $\sigma_x$ operator is applied to a control qubit, the controlled $\sigma_x$ gate will create an additional $\sigma_x$ operator on the target qubit as shown in Tab.~\ref{tab:CNOT}. This leads to the conclusion that for classical structures there is only one direction in which perturbations can spread. 
This direction is denoted in the network component scheme in the inset of Fig.~\ref{fig:ics} by black arrows. 
Only nodes that receive their input directly or indirectly from the initially perturbed node can exhibit perturbations.

The introduction of the Hadamard operation into our system allows for the perturbations to spread in both directions, allowing for a higher Hamming distance to arise in the quantum system. An example is shown in Fig.~\ref{fig:ics}. Each row shows the perturbation pattern for a different initial perturbation. As we can see, the classical network only allows the perturbations to spread alongside the arrows indicating the wiring direction (Fig.~\ref{fig:ics} top left). The perturbation in the quantum case is confined to the same portion of the network.

While the presence of a branch point in the network gives rise to complicated perturbation patterns in the quantum networks, the implementation of the logic in our $2N$ qubit systems consists only of linear operations.  This means, for example, that a collision between solitary states traveling in opposite directions cannot  produce complicated patterns; the two solitary states simply pass through each other.  The complexity generated by a branch point is a combinatorial effect associated with the timings of solitary states reflecting off of branch endpoints or traversing a loop to return to the branch point.

\section{Magnitude of classical and quantum Hamming distance} \label{sec:hammingstats}
In this section, we consider the total Hamming distance and its dependence on network size. Again, only $K=1$ (in-degree 1) networks are considered.  We construct ensembles of networks where the input to each node is randomly selected from the full set of nodes.  These ensembles thus contain networks consisting of several independent components, and the statistics of perturbation growth are heavily influenced by the statistics of sizes of individual components. 

\begin{figure}[t]
  \includegraphics[width=\linewidth]{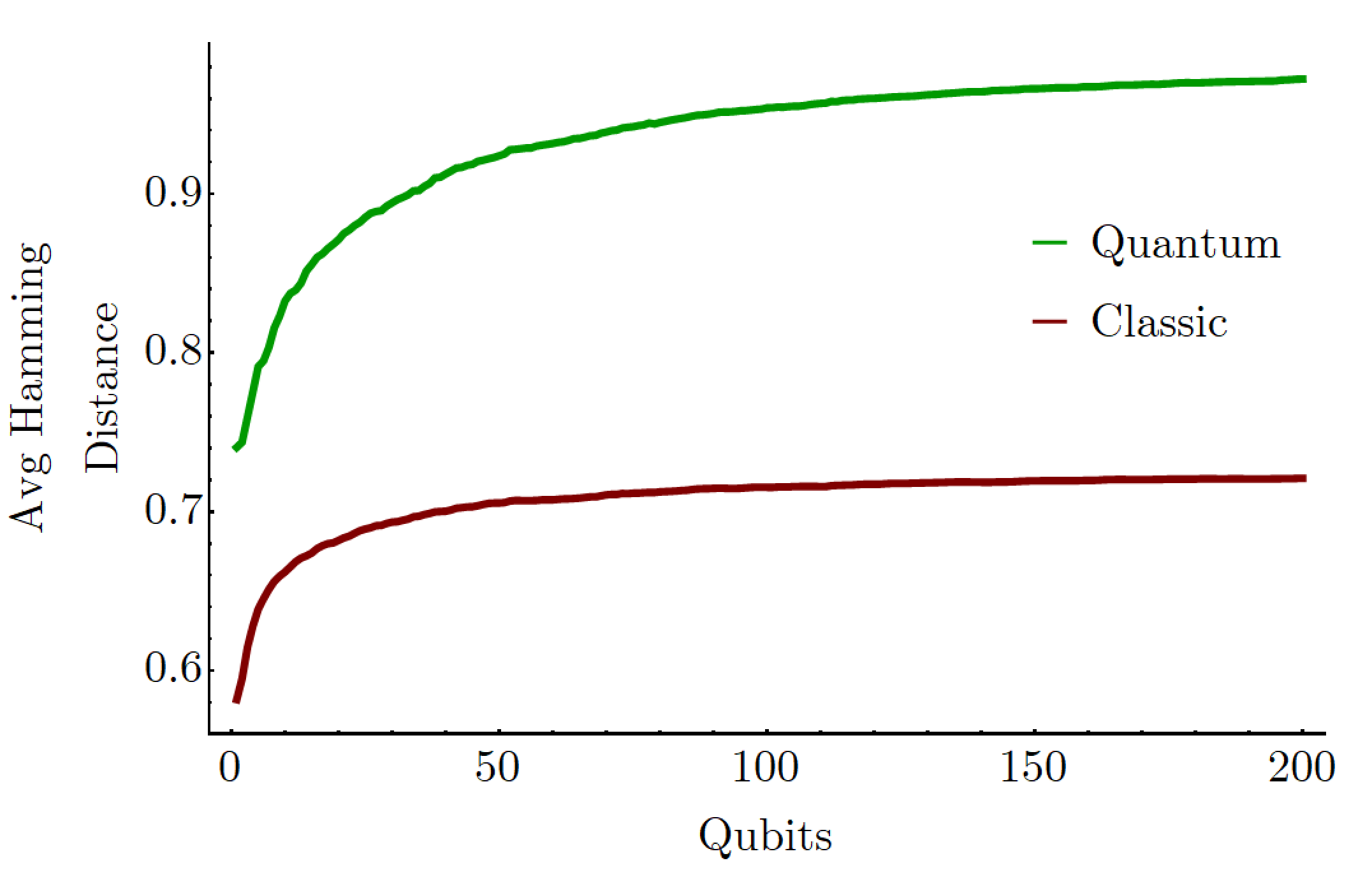}
  \caption{Averaged Hamming distance versus network size. For each network size an average over 1000 realizations over a period of 200 time steps is calculated. The initial perturbation locations are chosen randomly. For the quantum systems one Hadamard operation is applied on each target qubit before applying the logic function (compare Fig.~\ref{fig:network}d).}
  \label{fig:HD}
\end{figure}

Figure~\ref{fig:HD} displays the Hamming distance averaged over 1000 random realizations for each  network size. It can be seen that the magnitude of the quantum Hamming distance exceeds the classical one but exhibits a similar monotonic increase. As pointed out in previous research~\cite{coppersmith2001reversibleII}, the total Hamming distance is dependent on the network components of the system. Because initial perturbation (Eq.~\eqref{eq:comp}) is applied to a single qubit, the Hamming distance evolution is determined by the dynamics within that component alone.

Although the quantum Hamming distances of simple chains and loops do not exceed the classical ones, the average quantum Hamming distance is larger because most components contain branch points. It is very rare for a large component to be a simple loop or chain.  As we have seen, components with branch points give rise to complicated perturbation patterns with significantly higher Hamming distances than those reached in the classical networks. 

\section{Conclusion}

We have presented an approach to constructing quantum circuits that are natural generalizations of reversible random Boolean networks. Our formalism has allowed us to supplement the classical Boolean logic operations with quantum operations. 
The complexity of the quantum networks depends strongly on their connectivity. While generic systems show quasiperiodic dynamics, a certain nontrivial class of networks of single-input gates shows strictly periodic dynamics. For the special class showing periodic dynamics, we have extended the notion of the Hamming distance between trajectories  as a measure to investigate quantitatively the difference patterns generated by small perturbations applied to a network state.  For networks where each node has a single input ($K=1$), these patterns can differ dramatically from their classical counterparts. 

We have found that the propagation of perturbations is essentially different in quantum networks. In contrast to classical networks where the perturbations always spread, in the quantum case we find localized solitary perturbations moving through the network step by step. In particular, for simple chains and loops, the perturbation propagates as a localized solitary disturbance at constant velocity. 

An open challenge is to develop measures to analyze and interpret the perturbation dynamics generated in quantum networks with multi-input logic gates.

{\bf Acknowledgements.} We thank Iman Marvian for educational discussions about quantum circuits during the early stages of this work. This work was supported by the Deutsche Forschungsgemeinschaft (DFG, German Research Foundation) - Project Nos. 163436311 - SFB 910, 440145547, and 308748074. LK thanks DAAD for a scholarship and acknowledges the hospitality of Duke University, NC.

\bibliographystyle{apsrev4-1} % Tell bibtex which bibliography style to use
\bibliography{bibliography} % Tell bibtex which .bib file to use (this one is some example file in TexLive's file tree)

\end{document}